# Modelling a network where the opinion of each unit varies according to a "majority ruling" of its neighbouring units

V.F. Kusmartsev[1] and F.V. Kusmartsev[2]

[1]Trinity College, Cambridge University, Cambridge, CB2 1TQ

[2]Department of Physics, Loughborough University, LE11 3TU

**Abstract:**

The complexity of human behaviour can lead to very unpredictable patterns in social activity and structure. Here we demonstrate the instability of a community network controlled by majority ruling, where an element adopts the most popular opinion of their peers. We modelled a community as a square lattice, and performed sequential "time step" numerical calculations upon each cell in parallel. Depending on the initial ratio of two opinions, the community can segregate either into separate "gangs", or get dominated by a single opinion. We also note that gangs are separated by neutral or confused groups of individuals, buffering the transition. The behaviours shown by this model can be comfortably applied to many other real life situations, such as neural or ecological networks.

## 1 Introduction:

The complexity and unpredictable nature of human behaviour creates very interesting and diverse interactions in society. Society consists of a multitude of factors in various communities, each with its own hierarchies and social structures. These many factors, such as ideologies, conventions, morals and environment, can lead to complex dynamics in the generation of a dominant opinion in a certain group of individuals, whether the opinion is for which political party a group should vote for, what football team they support, and even more fundamental which signals may be sent in neural network within our brains.

Gabriel Tarde was the first who indicated that public opinion forms through interpersonal influence and conformity [1]. This issue is especially valid today due to the proliferation of online social media and chat networks, where individuals can spread or receive varying influences by their numerous contacts geographically spread over the whole world. Such various dynamics have been intensively studied in sociology [2,3,4] with the use of physics-like models and concepts [3,5,6] and use of statistical physics methods [7]. This approach to social networks was in general named "sociophysics".

There are many ongoing discussions in the world, where various groups are trying to convey disputable opinions to the overall public, whether they are supporting or are against contentious topics, such as nuclear energy, genetic engineering on living organisms, climate change and newly developed medicines. Each individual may or may not adopt new states in behaviour and opinion through the influence of their peers and neighbours. In general a consensus can be formed by an adoption of competing ideologies, traditions, and attitudes [8,2,3,9]. There have been original approaches to this dynamic opinion formation by Granovetter (the threshold model) [3] and by Bass (the diffusion of opinions) [10]. In both these models the key feature is that once an individual adopts a new state, his state remains unchanged at all subsequent times. In another approach, the opinion dynamics is described by the class of voter models [11,12]. In this approach, every node in the system may have one of two possible opinions, and at each time step a random nodes then copies the state of one or more of its randomly selected neighbours.

However, these previous models are less suited to studying the chaotic dynamics of competing opinions in situations where the opinion is likely to switch multiple times and in an unpredictable sequence until an overall consensus is reached. Such behaviour can in general be described with the use of a type of Ising spin model [13,14,15], which have been studied using statistical physics methods. There an individual's opinion is represented as a spin state, which can then be modelled as ferromagnetic ordering of spins in the Ising model associated with the ground state [16,13]. In many realistic situations the opinion-flips, during the evolution of a system, happen unconsciously [17] and in a very unpredictable chaotic way. However, in these spin flip processes, the dynamical, chaotic aspects have not been properly addressed.

Social communities and networks consist of a large number of socially interacting agents, and so can be compared to many-body systems in physics. The whole or part of such systems may exhibit macroscopic collective properties very similar to many situations arising in traditional statistical mechanics or physics. However the forces involved in social interactions which drive social dynamics are quite different in nature from forces existing in physical systems, and as such their mathematical description is far from complete. Traditionally, it is believed that social dynamics are driven by influence [3,7], homophily [18], attitudes and cognitive structural organization [20,21], social balance [22], reciprocity and topological network structure [23,24]. Recently a series of binary network models have been introduced, where the chaotic dynamics and decision making processes have been described with the aid of equations describing the temporal states of individual Ising spins [26,27,28]. Such an approach allows consideration of the many interacting and inter-coupled networks, including multiplex networks, that had also been introduced recently [29,30,31,32].

The social infrastructures of any community can be represented as a network of interacting elements with varying connectivity structures. Here we will consider a system that will contain 10,000 units that will adapt according to "majority ruling", where a unit will adopt the most dominant opinion amongst its neighbours.

## 2 Model and Numerical Experiments:

For this experiment a functional array of 100x100 will be used, with each element having an input from 4 neighbouring elements: above, below, left and right (see Figure 1). Each element can take one of three opinions: A, B or undecided, corresponding to the numbers 1, -1 and 0. Each element can be thought of as the variable $x_{ij}$, which is calculated according to the equation:

$$x_{ij}(t+1) = \text{sgn}\big(x_{i+1j}(t) + x_{i-1j}(t) + x_{ij+1}(t) + x_{ij-1}(t)\big) \qquad (1)$$

where the sum $(x_{i+1j} + x_{i-1j} + x_{ij+1} + x_{ij-1})$ is taken from the neighbouring four sites of the square lattices (see Figure 1).

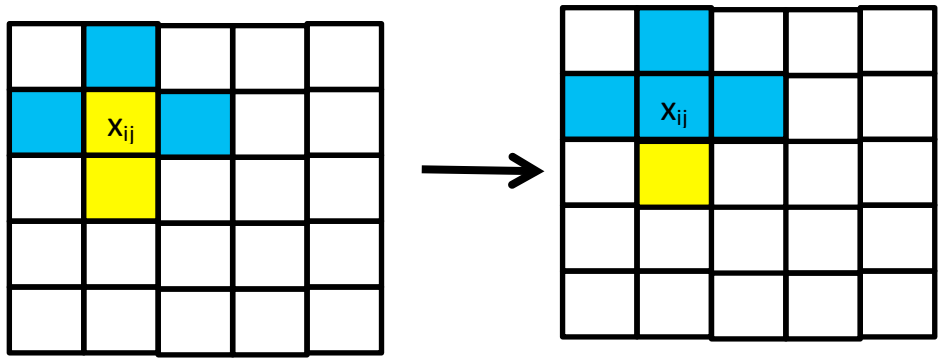

*Figure 1: A schematic illustration of equation (1). Opinion A is coloured blue and associated with $x_{ij}=1$, opinion B is coloured yellow ($x_{ij}=-1$). Here you can see $x_{ij}$ is surrounded by a majority of A opinions, and so after a time step iteration it too takes up the A opinion. Note $x_{ij}=0$ corresponds to an undecided individual, say colourless in this figure; and only the labelled cell has had the iteration applied to it.*

A set of scripts were created in python, using the modules numpy, random and mathplotlib in addition to the already present functions. The basis for each script was a square array that is filled randomly with either a random or known proportion (dependant on the desired area of analysis) of 1s and -1s. To simplify the calculations for each time step iteration, a border of 0s was left around the outside of the array. In these experiments, an array of 102x102 was used, leaving a functional area of 100x100 units (disregarding the 0 border).

A function, called iterate_1(Z,n), was created that would for each cell in the array sum the surrounding cells (4 in total), and then alter the value after that iteration is fully complete. If the sum is greater than 0, the cell becomes 1; if the sum is less than 0, the cell becomes -1, irrespective of what it was before. This form of iteration is known as parallel calculation, and was used for these experiments in order to remove bias of the various cells in the array. If a cyclic iteration method had been used, the cells appearing at the start of each cycle would have a greater effect on the final distribution, creating skew in the final results. In order to achieve parallel iterations, this function creates a new zero array N, fills it with the correct values collected from the original Z array, returning the new N array. To repeat the iteration until a constant final distribution is achieved, the original Z array is overwritten by the new N array, and this is repeated at least 20 times (value obtained by following the change in quantity of 1s and -1s in the array with increasing number of iterations. 20 iterations is always enough to reach a constant final distribution).

This iteration procedure was repeated for 1000 different randomly generated arrays of known, set initial conditions, and the final distributions were collected in separate data lists for the total sum of 1s and -1s in the final arrays. This data was then used to produce histograms and colourmaps to illustrate the results. The experiment was done firstly with every cell containing a value other than 0, and again including 0s as half of the array, in the initial set conditions.

# 3  Results:

First, we wanted to see how the array would change with each time step iteration. So using a script to collect the total sum of the 1s, -1s and 0s after each iteration and then plotting these values on a scatter graph, Figure 2 was obtained. We used the 100x100 array that we would then be using for the rest of the experiment. The "A votes" correspond to 1s in the array, "B votes" to -1s, and 0s correspond to "undecided" elements. As can be seen in Figure 2, the array reaches a final consensus fairly rapidly (after 15 iterations), which at first progresses in a linear fashion. We then decided to investigate whether increasing the number of 0s in the initial array would affect the rate of convergence to a consensus in the array. Hence we made an initial array with 5000 zeroes (50% of the array area), and followed its behaviour with each time step iteration, shown in Figure 3.

As can be seen in Figure 3, the array still converges to a solid consensus in roughly 15 time steps, and so this implies that the rate of convergence of our arrays according to the majority ruling is independent of the initial proportion of zeroes in the array. Due to this, we will be using at least 20 time step iterations to obtain the final distributions of each array in the later parts of this experiment. Also, in both Figure 2 and Figure 3, once the consensus had been reached, a slight fluctuation in the final numbers of As and Bs can be seen by the horizontal zigzag lines in the plots. This phenomenon is due to "confused" regions where the opinions of each element will switch continuously due to each element being surrounded by the opposite opinion. This will be visualised better in Figures 4 through 8 by the chessboard patterns (explained later).

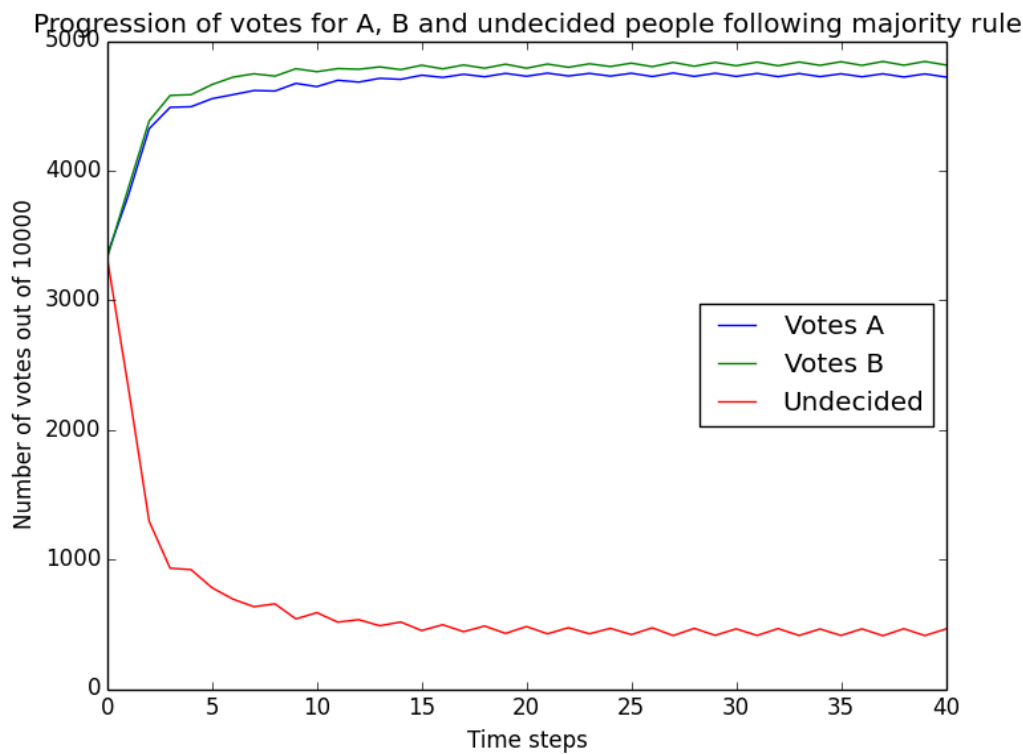

*Figure 2: An example scatter line plot showing how the total amount of each value(1=A, -1=B,0=undecided) varies as the number of iterations increases. The total number of cells was 10,000, aligned in a 100x100 array. The end zigzag pattern is due to continual switching of opinion*

*after each iteration, which can be seen as a "confused" region in the later colourmap. The initial conditions were random in this example.*

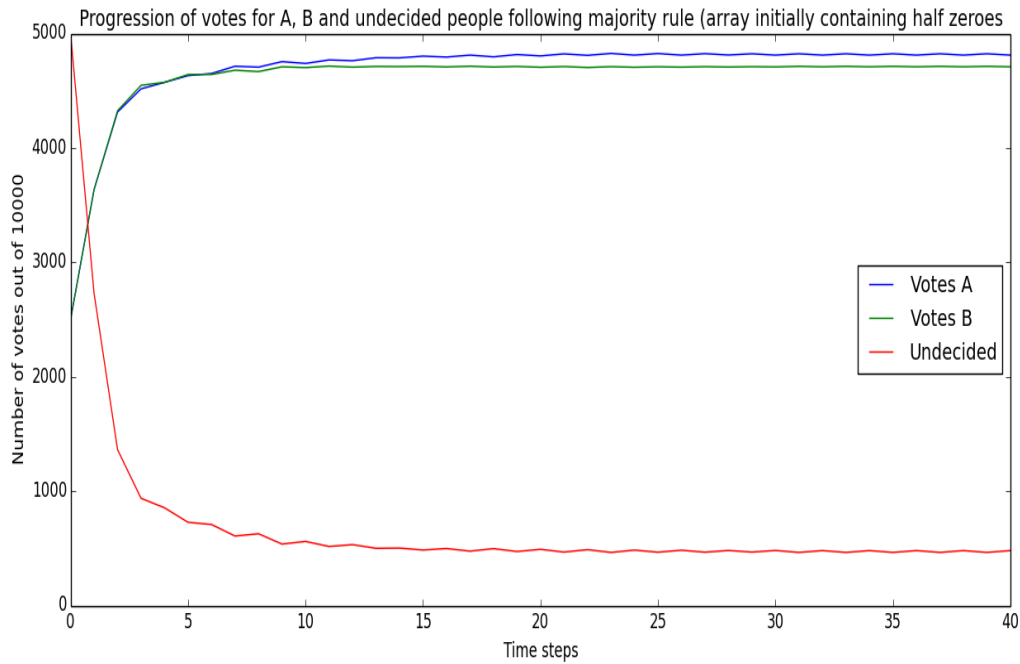

*Figure 3: A scatter line plot of an initial array with 50% of the initial elements as 0s (undecided). The rate of convergence to a consensus remains the same as in Figure 2, showing that the amount of zeroes in the initial array has a negligible effect on this convergence.*

We then started to lock the initial conditions of each array by controlling the initial numbers of 1s, -1s and 0s in each generated array. Note the values are still distributed randomly throughout the array, although the total number of each value is fixed. We then visualised the final consensus distribution of each array as colourmaps, where magenta elements correspond to 1s (or As), turquoise to -1s (or Bs), and lilac to 0s (or undecided) elements. The final arrays used to generate the colourmaps had 20 time step iterations applied to them. The number of the latter lilac elements is very small, implying that eventually the vast majority of elements in the array will adopt some opinion, irrespective whether they had an opinion initially or not. This is further highlighted comparing Figure 4 and Figure 5. Both these colourmaps where created from arrays where the initial A/B ratio was 1.0, but Figure 4 had 5000 zeroes initially, whereas Figure 5 had no zeroes. After 20 time step iterations, the final array produced in each cases show an almost identical pattern of opinion distribution. Both arrays have distinct "A groups" and "B groups" of almost identical sizes distributed randomly across the board. There are also clearly visualised "confused regions" as mentioned from Figures 2 and 3, which can be seen as the chessboard pattern of turquoise and magenta square elements in all the Figures 4 through 8. It is also seen that the majority of residual lilac undecided elements are found at the borders between these "confused regions" and other strongly opinionated groups. It is also quite rare that A and B groups form a smooth boundary between each other, without forming a confused region or having undecided elements buffering the transition. This is quite realistic, since often in life two groups of strongly differing opinions rarely coexist in harmony.

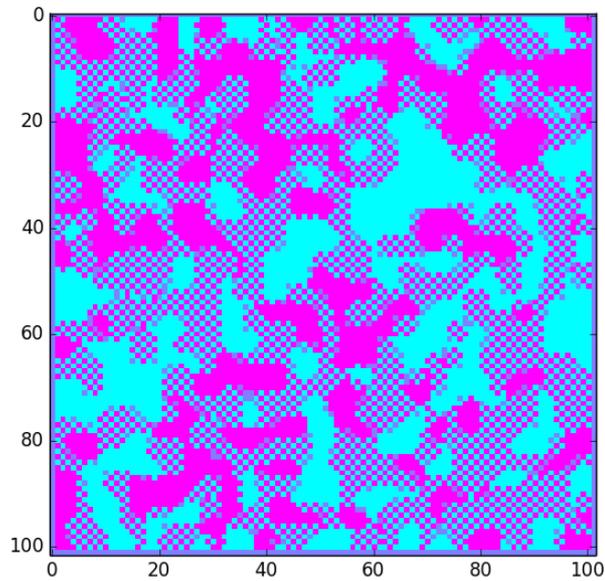

*Figure 4: A colourmap where the initial conditions had a A/B ratio of 1.0 and half the cells were undecided (i.e. 0s). The turquoise areas signify "B" opinions, the magenta areas signify "A" opinions, and lilac squares are undecided (the number of these is very small). The "confused" regions mentioned in figure 2 are the chessboard patterned sections, since their opinion will keep switching due to being surrounded by the opposite opinion.*

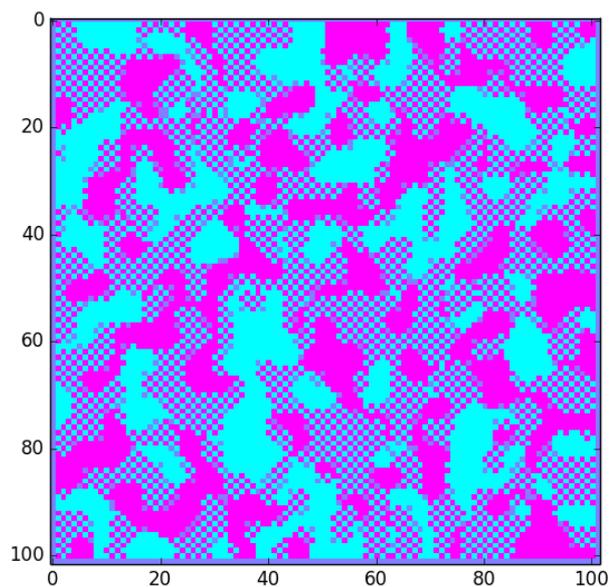

*Figure 5: A colourmap where the initial A/B ratio was 1.0, and every cell had been given an initial opinion. The overall distribution is very similar to the result obtained in figure 4.*

We then reduced the initial A/B ratio to 0.75 (Figure 6) and 0.50 (Figure 7) to see how that would affect the final distribution of the array after 20 time step iterations. As you can see in Figure 6, the slight reduction in the A/B ratio from 1.00 to 0.75 greatly reduces the final amount of A groups seen in the colourmap. There are only a few solid magenta A groups left in a vast sea of B opinion elements. Also, even in Figure 6 and especially in Figure 7, the majority of the least popular opinion actually occurs in confused regions, which is interesting since in this case, the magenta A opinion is still very unstable and can easily be switched to the more popular B opinion in these confused regions. In Figure 7 with an A/B ratio of 0.50, the A groups can be said to completely die out, with only a few residual A opinion elements floating around in the sea of B elements.

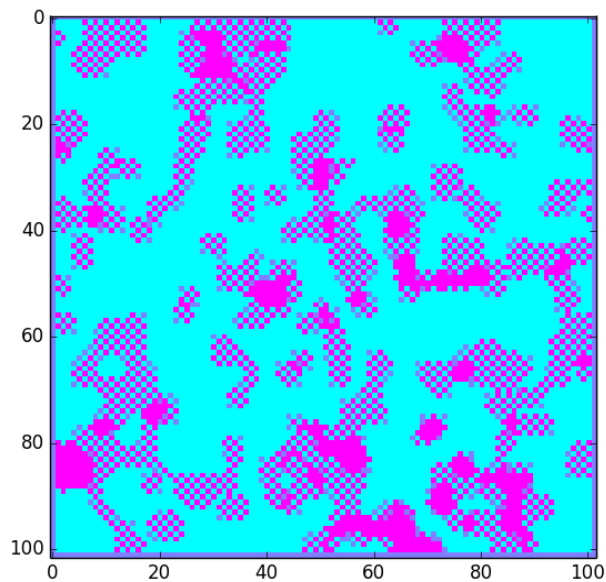

*Figure 6(above): A colourmap with initial A/B ratio of 0.75. Here the turquoise "B" groups have won over the majority of the array area, and there are very little so called magenta "A gangs", with most of the "A" opinion cells actually lying in confused regions.*

*Figure 7(below): A colourmap with initial A/B ratio of 0.5 leads to almost non-existant number of magenta "A gangs", and the vast majority conforming to the opinion B.*

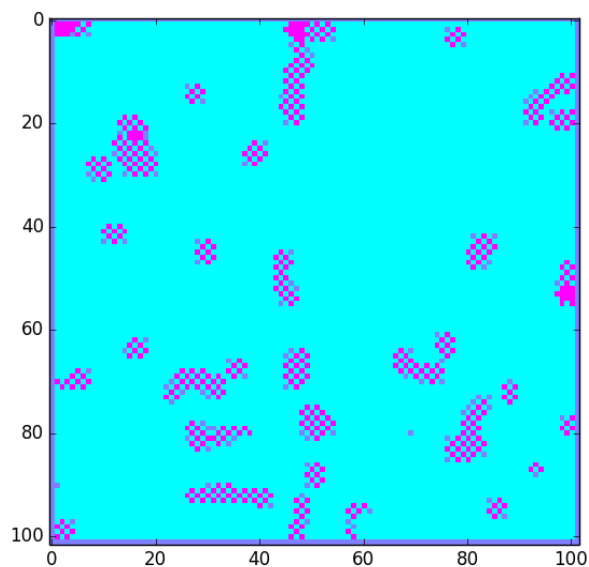

Due to the large difference that occurs between an initial A/B ratio of 1.00 and 0.75, colourmaps were generated for ratios 0.9 (Figure 8) and 0.8 (Figure 9a and b) to visualise the behaviour between these points. In Figure 8 where an A/B ratio of 0.90 was used, the overall structure of the colourmap is similar to Figure 5 and 6, with strong A and B groups dispersed across the map. The turquoise B groups however do start to join up and surround the smaller amount of magenta A groups. Going from a ratio of 0.90 to 0.80 seems to create a larger discrepancy between the amount of turquoise groups to magenta groups, leading more to the structure of magenta "A" islands in a turquoise sea of "B" gangs.

Another important result that is seen comparing Figures 9a and 9b is that by having half the array initially filled with zeroes, the discrepancy between the amount of A to the amount of B is reduced. This implies that undecided elements can reduce and weaken the strength of progression of a dominant opinion as it attempts to conquer territory with consequent time step iterations. This shows, that even though the zeroes didn't have any effect on the rate of convergence, as shown in Figures 2 and 3, the amount of zeroes can affect the strength of dominance exhibited by the greater opinion over the whole array space.

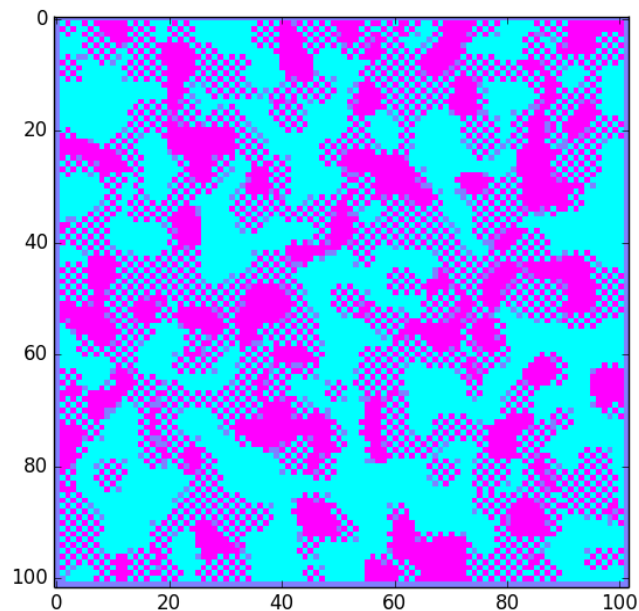

*Figure 8: A colourmap with initial A/B ratio of 0.90. This shares a similar structure with Figures 4 and 5 with separate clumps of turquoise and magenta. The turquoise here is only starting to dominate the array space.*

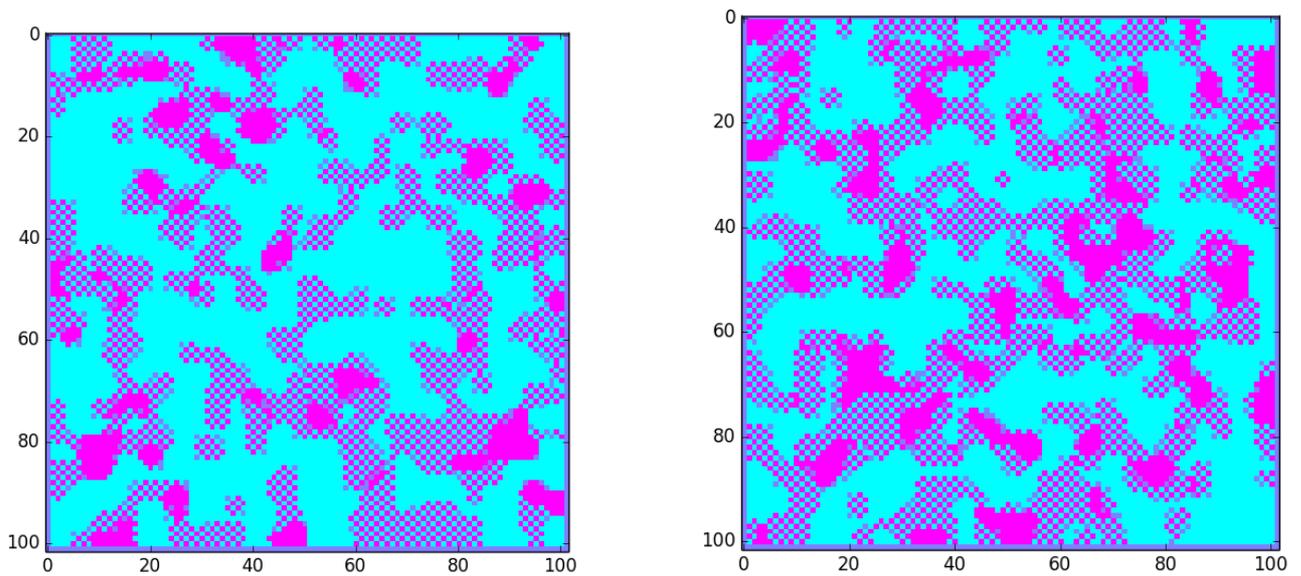

*Figure 9a(left) and b(right): 9a is a colourmap of initial ratio 0.80 and every element having an initial opinion. Here the turquoise B opinion has already conquered the great majority of array territory, leaving only a few magenta A groups. 9b has the same initial ratio, however half the initial elements are zeroes. This shows the initial zeroes lessen the strength of territory dominance exhibited by the majority opinion (B in this case).*

In order to further analyse these results, histograms were produced using data of the final distributions obtained from 1000 different randomly generated arrays. In a single such calculation, all the 1000 arrays had the initial A/B ratios, whether every element had a value or not, but where these values were located were assigned randomly. The histograms allow us to see the results of many repeated calculations, and so obtain the average and deviations of these array calculations. 40 time step iterations were used to remove the chance of unwanted variation in the final proportions of each array, since only the distributions of the final consensae want to be collected and compared.

As can be seen in Figures 10 and 11, for an initial ratio of 1.0, the amount of A and B cells form overlapping normal distributions with mean of about 4800, the mean for Figure 11 being slightly lower due to the large amount of initial zeroes seeming to increase the final number of zeroes as well. We then decreased the initial A/B ratio down to 0.75 which was the other extreme (seen in Figure 6), where the B opinion greatly dominated the final consensus. This produced Figures 12 and 13, where it can be seen that B voters vastly outnumber A voters. The standard deviations of each group of peaks seem to stay constant, whereas the means for both the A and B move almost equally in either direction. The calculation including 50% zeroes initially produces Figure 13, which shows the separation between the numbers of As and Bs is reduced, which was also seen qualitatively in Figures 9a and b. This leads to the question whether the zeroes merely delay the separation in amounts of each opinion, or reduce the magnitude of separation linearly. It is also interesting to find the point where the arrays start to produce a singular dominant opinion. For this reason, the calculations were repeated with initial ratio of 0.90, with and without initial zeroes. These results are shown in Figures 14 and 15.

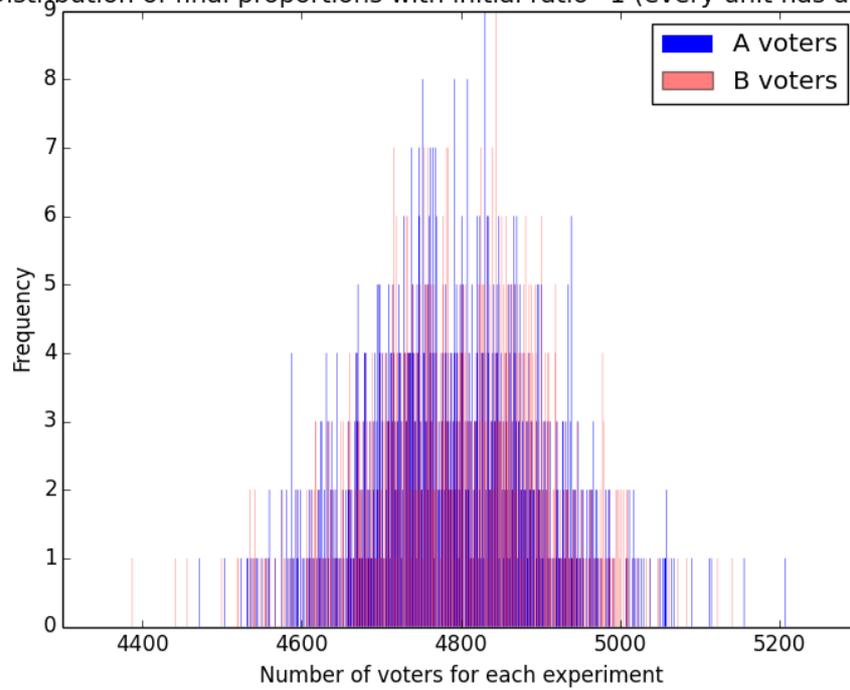

*Figure 10: A histogram of the final amounts of A and B cells in an array after 40 iterations. 1000 independent arrays had been randomly generated with known initial conditions of ratio=1.0 and no empty cells, with the final values collected and visualised in this histogram. 1000 bins were use. Both the numbers of As and Bs form normal distributions with mean approximately 4800.*

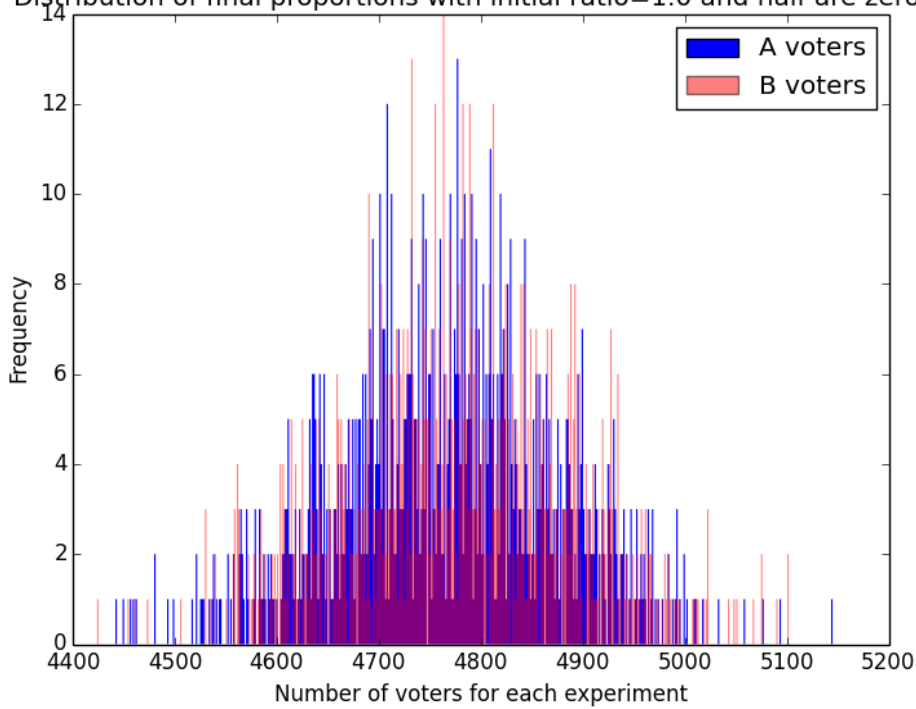

*Figure 11: A histogram similar to Figure 10 apart from initially half the cells are empty. The mean of both normal distributions is slightly lower (approx. 4770), implying there are slightly more zeroes in the final array, but still a negligible amount in respect to the whole array.*

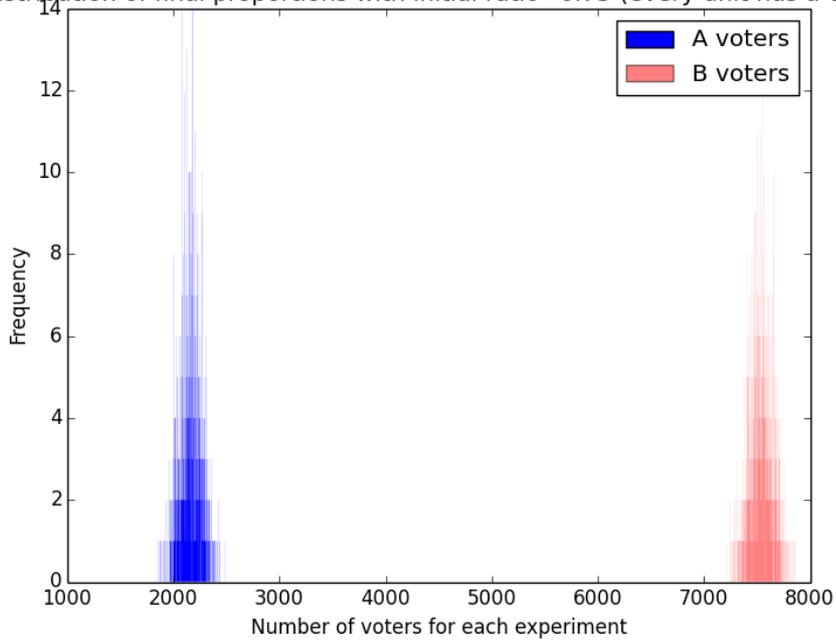

*Figures 12(above), 13(below): Histograms produced via the same method as Figures 10 and 11, but with initial ratio A/B = 0.75. Figure 12 has no zeroes initially, whereas Figure 13 has 50% zeroes initially. Here a drastic separation in the number of As and Bs is seen, a larger separation visible in Figure 12 than Figure 13. Both parts keep the normal distribution shape.*

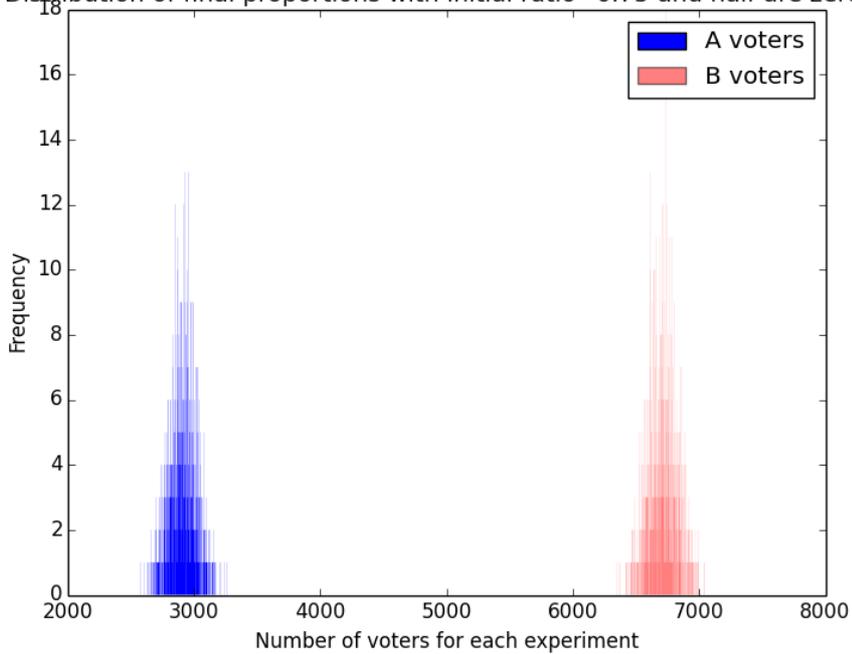

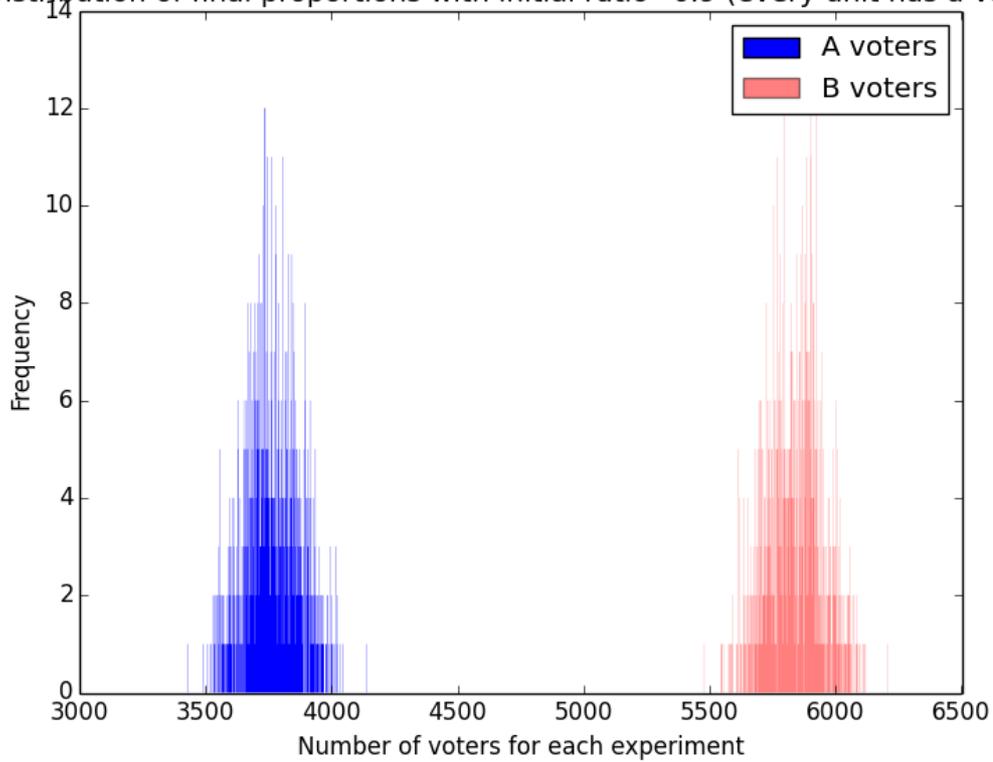

*Figure 14 (above) and 15 (below): Same as the figure 12 and 13 pairing, except with initial A/B ratio 0.9. Here the separation is still clear, but of a much smaller magnitude. The standard deviations of each peak remain the same as in Figures 10 and 11.*

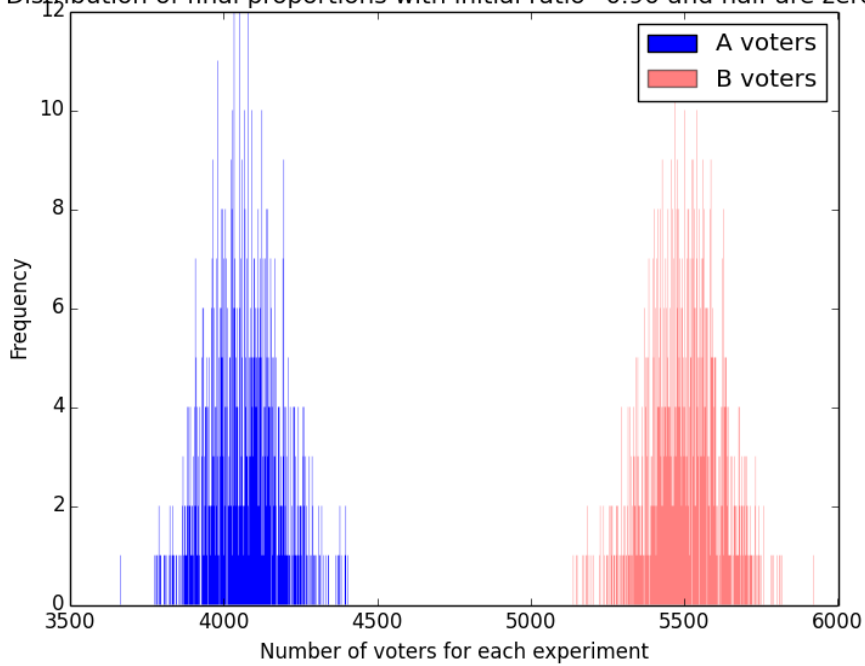

As can be seen comparing Figure 14 and 15, the separation between the A and B peaks is smaller with the initial zeroes, yet there is still a clear separation. This suggests that the zeroes do not affect when the peaks begin to separate, and just affect the rate at which they will spread. This could be explained by considering the initial zeroes as a sort of buffer, weakening the strength of initial opinionated gangs and preventing a rapid and large change in array distribution. Also, the zeroes strengthen the minority opinion, by reducing the number of bordering elements required to change the cell from 3 to 2 (3/4 vs 2/3).

## 4 Conclusions and Discussion:

The most striking result seen in both the colourmaps and the histograms is that a very small change in A:B ratio away from 1:1 will lead to a significant discrepancy in the final numbers of each opinion in the network. Also, as shown in Figure 2, the speed with which the model will arrive at its final distribution is fairly rapid and linear, and independent of the initial amount of zeroes in the array. The prevailing opinion will dominate a network irrespective of its initial distribution, as shown by the separation of the normal distribution peaks in Figures 10 through 15. However, the greater the number of zeroes in the initial array, the weaker this separation of opinions is, even though there are very few undecided elements in the final consensus array.

There are multiple examples which this model can help to simplify and explain. For example, in neural networks, summation synapses are used very often to modulate sensory signals and control output to and from the control centres in the brain. Depending whether a hyperpolarising or depolarising action potential is prevalent ultimately determines what signal is sent across the network. The only difference in this simplified example is that in this model our threshold is 0, whereas the threshold in a neuron is set around +40mV.

Another example is social networks, where a group of contacts may discuss their opinions for either supporting a football team in the upcoming world cup, or what political party they would like to vote for in the elections. However there are multiple assumptions made by this model: One being that people will change their opinion readily only dependant on their friends opinions, another that the persons initial opinion has no weight in what their new opinion will be after conversing with friends (ie after an iteration). Also, in real life a lot of people will listen to certain friends more than their other friends, and so the strengths of these network links would be higher, whereas in this model the strengths have all been simplified to 1. An improvement that could be done on this model would be to add the weights of links, converting it more to the McCulloch-Pitts model (Figure 16). This change would also make the model more valid in respect to simulating neural networks as well.

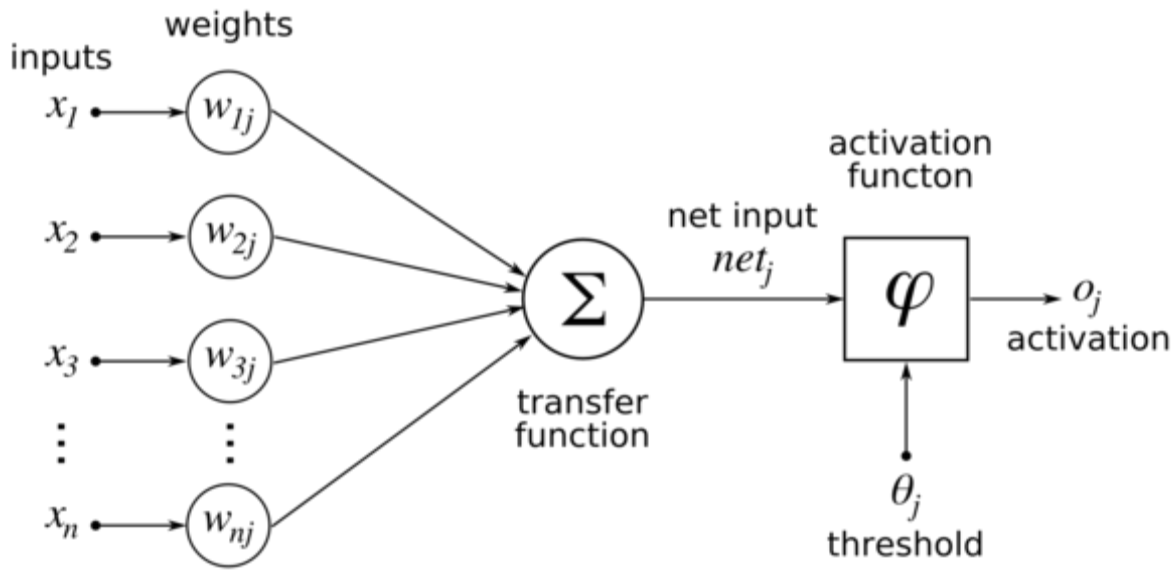

*Figure 16: Diagram of the McCulloch-Pitts model for neural networks (source en.wikibooks.org). Each input has a different weight to the final sum, which then determines the output signal. This can also be used in social networks as an improvement on the majority model made in this experiment.*

Useful comparisons can be made to the rules of the "game of life" model by Jon Conway [33], where the ruling leads to a cell staying alive only when there is a specific number of allies nearby (2 or 3 in this case), which can lead to very interesting patterns emerging, such as the "glider". The simpler rules in this majority ruled model leads to a much faster convergence rates, and does not allow new opinions to form inside an already established "gang".

Another possible use of the majority model could be in determining the dynamics of two species fighting for a singular ecological niche. These two species have the same food sources, the same predators and live in the same habitats, so only one can exist in any one cell. If a niche is surrounded by species A, then that species will spread into that cell, outcompeting any species B that may have already been set up there. An undecided cell could be the result of the two species being equal in strength, leaving a wasteland where neither species can populate that element. This example will behave very similarly to the majority rules created in this experiment.

Thus, in this experiment we have managed to explain and highlight key behaviours and complex structures of a fairly simple set of rules, where the majority prevails. The results of this work have been published in the Ref. [34].